\documentclass[prb,superscriptaddress,twocolumn]{revtex4-1}
\usepackage[utf8]{inputenc}
\usepackage{amsmath}
\usepackage{amsfonts}
\usepackage{amssymb}
\usepackage{graphicx}
\usepackage{lipsum}
\usepackage{hyperref}
\usepackage{xcolor}
\usepackage{multirow}

\begin{document}

\title{Dzyaloshinskii-Moriya interaction and the magnetic ground state in magnetoelectric LiCoPO$_4$}
\date{\today}

\author{Ellen Fogh}
\affiliation{Department of Physics, Technical University of Denmark, DK-2800 Kongens Lyngby, Denmark}
\author{Oksana Zaharko}
\affiliation{Laboratory for Neutron Scattering and Imaging, Paul Scherrer Institute, Villigen CH-5232, Switzerland}
\author{Jürg Schefer}
\affiliation{Laboratory for Neutron Scattering and Imaging, Paul Scherrer Institute, Villigen CH-5232, Switzerland}
\author{Christof Niedermayer}
\affiliation{Laboratory for Neutron Scattering and Imaging, Paul Scherrer Institute, Villigen CH-5232, Switzerland}
\author{Sonja Holm-Dahlin}
\affiliation{Laboratory for Neutron Scattering and Imaging, Paul Scherrer Institute, Villigen CH-5232, Switzerland}
\affiliation{Nano-Science Center, Niels Bohr Institute, University of Copenhagen, DK-2100 Copenhagen Ø, Denmark}
\author{Michael Korning Sørensen}
\affiliation{Department of Physics, Technical University of Denmark, DK-2800 Kongens Lyngby, Denmark}
\author{Andreas Bott Kristensen}
\affiliation{Department of Physics, Technical University of Denmark, DK-2800 Kongens Lyngby, Denmark}
\author{Niels Hessel Andersen}
\affiliation{Department of Physics, Technical University of Denmark, DK-2800 Kongens Lyngby, Denmark}
\author{David Vaknin}
\affiliation{Ames Laboratory and Department of Physics and Astronomy, Iowa State University, Ames, Iowa 50011}
\author{Niels Bech Christensen}
\affiliation{Department of Physics, Technical University of Denmark, DK-2800 Kongens Lyngby, Denmark}
\author{Rasmus Toft-Petersen}
\affiliation{Department of Physics, Technical University of Denmark, DK-2800 Kongens Lyngby, Denmark}

\begin{abstract}

Magnetic structures are investigated by means of neutron diffraction to shine light on the intricate details which are believed key to understanding the magnetoelectric effect in LiCoPO$_4$. At zero field, a spontaneous spin canting of $\varphi = 7(1)^{\circ}$ is found. The spins tilt away from the easy $b$-axis towards $c$. Symmetry considerations lead to the magnetic point group $m'_z$ which is consistent with the previously observed magnetoelectric tensor form and weak ferromagnetic moment along $b$. For magnetic fields applied along $a$, the induced ferromagnetic moment couples via the Dzyaloshinskii-Moriya interaction to yield an additional field-induced spin canting. An upper limit to the size of the interaction is estimated from the canting angle.

\end{abstract}

\maketitle

\section{Introduction}

In a number of insulators, an external electric or magnetic field can induce a finite magnetization or electric polarization respectively. This so-called magnetoelectric (ME) effect was first theoretically predicted \cite{landau_lifshitz,dzyaloshinskii1959} and shortly thereafter experimentally observed in Cr$_2$O$_3$ \cite{astrov1960,astrov1961}. Since then, a collection of materials displaying the ME effect has been identified but the underlying microscopic mechanisms are not yet fully understood.

The Dzyaloshinskii-Moriya (DM) interaction has proved a key ingredient in explaning the induced or spontaneous electric polarization in a number of compounds such as $R$MnO$_3$ ($R$ = Gd, Tb, Dy) \cite{sergienko2006}, Ni$_3$V$_2$O$_8$\cite{kenzelmann2006} and CuFeO$_2$\cite{kimura2006}. In these systems, the non-collinear incommensurate order of the magnetic moments results in a displacement of the oxygen ions situated in between neighboring moments and a net displacement of charge is generated\cite{kimura2007}. Non-collinear order may appear as a consequence of competing interactions, so-called spin frustration. Such systems are associated with large ME effects\cite{kimura2007,spaldin2008}. 

The lithium orthophosphate family (space group \textit{Pnma}), Li$M$PO$_4$ ($M$ = Co, Ni, Mn, Fe), is in many ways an excellent model system for studying the ME effect. All family members exhibit commensurate near-collinear antiferromagnetic order as well as the ME effect in their low-temperature and low-field ground state. In recent studies, additional ME phases were found at elevated magnetic fields applied along the respective easy axes in LiNiPO$_4$ \cite{toftpetersen2017} and LiCoPO$_4$ \cite{khrustalyov2016}. In both materials, these high-field ME phases are also accompanied by commensurate antiferromagnetic order \cite{fogh2017,toftpetersen2017}.

The magnetically induced linear ME coupling is described as $P_i = \alpha_{ij} H_j$, where $P_i$ is the electric polarization, $H_j$ the external magnetic field and $\alpha_{ij}$ are the ME tensor elements with $i,j = \lbrace a,b,c \rbrace$. Allowed tensor elements are dictated by the magnetic symmetry of the system. For collinear (anti)ferromagnets one may think of tensor elements for which the magnetic field is \textit{perpendicular} to the spin orientation, $\alpha_{\perp}$, and those for which the field is \textit{parallel} to the spins, $\alpha_{\parallel}$. Magnitudes and temperature dependencies for $\alpha_{\perp}$ and $\alpha_{\parallel}$ have been computed from first principles for ME compounds such as Cr$_2$O$_3$ \cite{iniguez2008,mostovoy2010,malashevich2012,mu2014}, and LiFePO$_4$ \cite{scaramucci2012}. In these studies it is possible to separate effects due to ion displacements within the unit cell (lattice contribution) and effects due to electronic motion around 'clamped' ions (electronic contribution). In both cases one distinguishes between spin and orbital effects.

\begin{figure}[t!]
	\centering
	\includegraphics[width = \columnwidth]{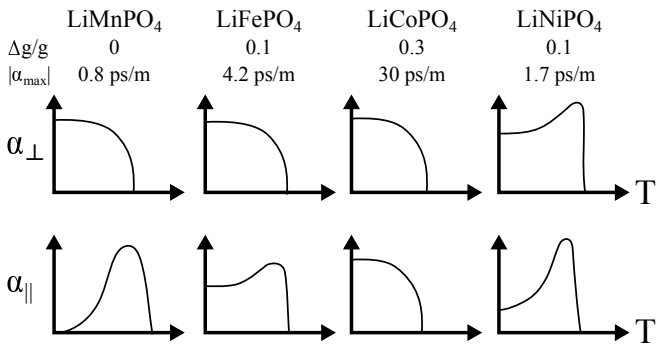}
	\caption{Schematic of temperature dependencies of the ME tensor elements, $\alpha_{\perp}$ and $\alpha_{\parallel}$, for the lithium orthophosphates as measured by Mercier \cite{mercier}. Values of $\Delta g/g$ \cite{toftpetersen2015} are listed for each compound as well as maximum absolute values of the ME coefficients \cite{rivera1994,wieglhofer}.}
	\label{fig:ME}
\end{figure}

The \textit{ab initio} calculations show that $\alpha_{\perp}$ is generally dominated by the spin-lattice contribution and the ME coupling is relativistic in origin, e.g. via the DM interaction. The predicted temperature dependence of $\alpha_{\perp}$ follows that of the order parameter \cite{malashevich2012,scaramucci2012}. This corresponds well with observations in the lithium orthophosphate family except for a slight variation in the curve for LiNiPO$_4$ [see Fig. \ref{fig:ME}].

The behavior of $\alpha_{\parallel}$ is altogether more tricky and \textit{ab initio} calculations indicate that orbital contributions may play an important role \cite{malashevich2012,scaramucci2012}. When disregarding orbital contributions, the computed temperature dependence of $\alpha_{\parallel}$ displays a maximum below the transition temperature and then goes to zero for $T \rightarrow 0$ \cite{mostovoy2010}. The comparison of the measured and predicted temperature dependencies of $\alpha_{\parallel}$ for LiMnPO$_4$ is excellent [see Fig. \ref{fig:ME}]. However, for the remaining family members $\alpha_{\parallel} \neq 0$ for $T \rightarrow 0$ and the prediction is clearly lacking. It is even worse in the case of Cr$_2$O$_3$ (not shown) where $\alpha_{\parallel}$ changes sign as a function of temperature \cite{rado1961}. The orbital moment is almost entirely quenched for LiMnPO$_4$ but not for LiFePO$_4$, LiCoPO$_4$ and LiNiPO$_4$. Hence, the discrepancy between the predicted and measured values of $\alpha_{\parallel}$ for $T \rightarrow 0$ for the latter three compounds may be related to the orbital moment. Moreover, the maximum magnitude of the observed ME tensor elements also appears linked to the orbital moment with $\Delta g/g = 0$ and $|\alpha_{\mathrm{max}}| = 0.8\,\mathrm{ps/m}$ for LiMnPO$_4$ and $\Delta g/g = 0.3$ and $|\alpha_{\mathrm{max}}| = 30\,\mathrm{ps/m}$ for LiCoPO$_4$. However, recent first-principle calculations on LiFePO$_4$ taking into account orbital contributions (both lattice and electronic parts) still fail to encapsulate the low-temperature behavior of $\alpha_{\parallel}$ \cite{scaramucci2012}.

In this paper, we focus on LiCoPO$_4$ which has, by far, the strongest ME effect in the lithium orthophosphate family \cite{mercier,wieglhofer}. Although intensively studied, there is as of yet no satisfactory theory for the underlying microscopic mechanism. LiCoPO$_4$ has lattice parameters $a = 10.20\,\mathrm{Å}$, $b = 5.92\,\mathrm{Å}$ and $c = 4.70\,\mathrm{Å}$ \cite{newnham1965} and the four magnetic Co$^{2+}$ ions ($S = \frac{3}{2}$) of the crystallographic unit cell form an almost face-centered structure with the positions ${\bf r}_1 = (1/4+\varepsilon, 1/4, 1-\delta)$, ${\bf r}_2 = (3/4+\varepsilon, 1/4, 1/2+\delta)$, ${\bf r}_3 = (3/4-\varepsilon, 3/4, \delta)$ and ${\bf r}_4 = (1/4-\varepsilon, 3/4, 1/2-\delta)$ and with the displacements $\varepsilon = 0.0286$ and $\delta = 0.0207$ \cite{kubel1994}. The zero-field commensurate antiferromagnetic structure of LiCoPO$_4$ has spins along $b$ (easy axis) and the four magnetic ions in a $C = (\uparrow \uparrow \downarrow \downarrow)$ arrangement \cite{santoro1966}. Here $\uparrow$/$\downarrow$ denotes spin up/down for ions on site number $1-4$. The transition temperature is $T_N = 21.6\,\mathrm{K}$ \cite{szewczyk2011,vaknin2002} and the saturation field is $\sim28\,\mathrm{T}$ with saturated moment $3.6\,\mathrm{\mu_B}$/ion \cite{kharchenko2010}. A number of studies establish that the magnetic point group of the zero-field magnetic structure is $2'_x$ rather than $mmm'$ as previously believed \cite{santoro1966}. This is based on the observation of a weak ferromagnetic moment \cite{rivera1994,kharchenko2003}, symmetry of the susceptibility tensor of optical second harmonic generation \cite{zimmermann2009,vanAken2008} and the discovery of a toroidal moment \cite{vaknin2002,ederer2007,spaldin2008,vanAken2007,zimmermann2014}. The magnetic phase diagram of LiCoPO$_4$ was previously characterized up to $25.9\,\mathrm{T}$ applied along $b$ by magnetization measurements, neutron diffraction and electric polarization measurements \cite{kharchenko2010,fogh2017,khrustalyov2017}. At $11.9\,\mathrm{T}$, the commensurate low-field structure gives way to an elliptic spin cycloid propagating along $b$ with a period of thrice the crystallographic unit cell. The magnetic moments are in the $(b,c)$-plane with the major axis along $b$. In the field interval $20.5-21.0\,\mathrm{T}$, the magnetic unit cell size remains but the spins re-orient. Above $21.0\,\mathrm{T}$, there is a re-entrance of commensurate magnetic order accompanied by the ME effect.

In this work we investigate the possible role of the spin-orbit coupling for explaining the ME effect in LiCoPO$_4$ as well as its sister compounds. In order to do so we look into the details of the zero-field magnetic structure of LiCoPO$_4$ and study the effects of a magnetic field applied along $a$ by means of neutron diffraction and magnetometry. A spontaneous canting of spins away from the $b$-axis towards $c$ is revealed. The resulting structure has magnetic point group $m'_z$ and we discuss the implications related to the ME tensor form and with regards to previous studies. In order to investigate the DM interaction in LiCoPO$_4$ we perform a neutron diffraction experiment with magnetic fields applied along $a$, i.e. perpendicular to the easy axis. The induced ferromagnetic moment couples via the DM interaction to yield a field-induced spin canting. We estimate the size of the DM interaction and discuss how this interaction may play a part as generator of the ME effect in LiCoPO$_4$.

\section{Experimental details}

Vibrating sample magnetization measurements were performed with a standard CRYOGENIC cryogen free measurement system. Magnetic fields of $0-16\,\mathrm{T}$ were applied along $a$ for temperatures in the interval $2-300\,\mathrm{K}$.

The zero-field magnetic structure was determined at the TriCS diffractometer at the Paul Scherrer Institute (PSI) employing an Euler cradle, a closed-cycle He refrigerator, open collimation and a Ge(311) monochromator with wavelength $\lambda = 1.18\,\mathrm{Å}$. No $\lambda/2$ contamination of the beam is possible due to the diamond structure of Ge. 193 inequivalent peaks were collected at $30\,\mathrm{K}$ and $5\,\mathrm{K}$.

Canting components of the zero-field structure could not be unambigiously determined at TriCS due to extinction effects and the large absorption cross section of Co. Instead, these components were investigated at the triple-axis spectrometer RITA-II at the PSI where a low background is obtained by energy discrimination. The instrument was operated in elastic mode with incoming and outgoing wavelength $\lambda = 4\,\mathrm{Å}$. A PG(002) monochromator and 80' collimation between monochromator and sample were used and a liquid nitrogen cooled Be filter after the sample ensured removal of $\lambda/2$ neutrons. A cryomagnet supplied vertical magnetic fields up to $12.2\,\mathrm{T}$ along $a$ and $b$ for samples oriented with scattering planes $(0,K,L)$ and $(H,0,L)$ respectively.

A high quality LiCoPO$_4$ single crystal measuring $2\times2\times5\,\mathrm{mm^3}$ ($\sim20\,\mathrm{mg}$) was used for magnetization measurements with magnetic fields applied along $a$ and for neutron diffraction experiments in zero field and with magnetic fields applied along $b$. A second sample with dimensions $3\times4\times4\,\mathrm{mm^3}$ ($\sim40\,\mathrm{mg}$) was used for the neutron diffraction experiment performed with fields applied along $a$.

\section{Results \& discussion}

The atomic and magnetic structures of LiCoPO$_4$ were determined by combining data from the TriCS and RITA-II experiments. Based on the \textit{Pnma} space group and 241 Bragg peaks, atomic displacements of $\varepsilon = 0.028$ and $\delta = 0.020$ were refined in Fullprof \cite{rodriguez-carvajal1993} ($R_F = 11.9$\%) in fair agreement with literature\cite{kubel1994}. The zero-field magnetic structure was determined from 130 Bragg peaks and is mainly of $C_y$ symmetry ($R_F = 17.2$\%), a result conforming with earlier findings\cite{santoro1966,vaknin2002}. The refined magnetic moment is $3.54(5)\,\mathrm{\mu_B}$, consistent with previous magnetization measurements \cite{kharchenko2010}. Note that the Li occupancy was refined to $1.03(5)$ and hence the sample is stoichiometric within the precision of the experiment. Refinement results with the Li occupancy fixed to unity are listed in Table \ref{tab:fullprof}.

\begin{table}[t!]
	\centering
	\caption{Atomic positions for LiCoPO$_4$ obtained from Fullprof refinement ($R_F = 11.9$\%) using 241 Bragg peaks collected at TriCS at $(30\,\mathrm{K}, 0\,\mathrm{T})$ and using the \textit{Pnma} space group. The Debye-Waller factor was fixed to $B_{\mathrm{iso}} = 0.20$. The magnetic moment in  units of $\mu_B$ as refined using a $C_y$ symmetry component is given in the rightmost column. This results from refinement ($R_F = 17.2$\%) using 130 commensurate magnetic peaks collected at $(2\,\mathrm{K},0\,\mathrm{T})$. The lattice parameters used in the refinements were $a = 10.20\,\mathrm{Å}$, $b = 5.92\,\mathrm{Å}$ and $c = 4.70\,\mathrm{Å}$ as given in Ref.\onlinecite{newnham1965}.}
	\label{tab:fullprof}
    \begin{ruledtabular}
    \begin{tabular}{c c c c c c}
    	Atom	& Site	& $x$		& $y$		& $z$			& $R_y$\\
    	\hline
	    	Li 		& 4a	& 0			& 0			& 0			& --\\
		Co 		& 4c	& 0.278(2) 	& 0.25		& 0.980(3)	& 3.54(5)\\
		P		& 4c	& 0.0945(8) 	& 0.25		& 0.419(2)	& --\\
		O1		& 4c	& 0.0986(7) 	& 0.25 		& 0.743(2)	& --\\
		O2		& 4c	& 0.4545(7) 	& 0.25 		& 0.203(1)	& --\\
		O3		& 8d	& 0.1669(5) 	& 0.0463(7)	& 0.2826(9)	& --	
    \end{tabular}
    \end{ruledtabular}
\end{table}

Other magnetic structures including a minor spin rotation towards $c$ ($C_z$) or a spin canting towards $c$ ($A_z$) were proposed but these refinements were not sufficiently different to distinguish them from the one regarding only a $C_y$ component. Extinction effects and the large neutron absorption cross section of cobalt result in significantly different intensities for equivalent Bragg peaks and hence, the TriCS data only enabled identification of the major symmetry component, $C_y$. 

Minor spin components in zero field and for magnetic fields applied along $b$ and $a$ were investigated at RITA-II by measuring a few key Bragg peaks: $(3,0,1)$, $(0,1,0)$, $(1,0,0)$, $(0,2,1)$, $(0,1,2)$ and $(0,0,1)$. Of these only $(0,1,0)$ has zero magnetic intensity. The calculated magnetic structure factors for the four basis vectors, $|S_R({\bf Q})|^2$, $R = \left\lbrace A, G, C, F \right\rbrace$, and spin polarization factors, $|P_i({\bf Q})|^2$, $i = \left\lbrace x,y,z \right\rbrace$, for these peaks are listed in Table \ref{tab:structurefactors}. The magnetic neutron intensity may then be expressed as:
\begin{equation}
	I({\bf Q}) \propto S^2 \, f({\bf Q})^2 \, \sum_{R} |S_R({\bf Q})|^2 \sum_{i} |P_i({\bf Q})|^2,
	\label{eq:intensity}
\end{equation}
where $f({\bf Q})$ is the magnetic form factor and $S$ is the thermal average of the magnetic moment. The following analysis is based on a process of eliminating possible structures and is not a full structure refinement.

\begin{table}[!b]
	\caption{Absolute squares of structure and polarization factors for the magnetic basis vectors reflected by the key Bragg peaks used to establish the magnetic structure of LiCoPO$_4$. The factors are normalized to unit spin lengths. Note that the crystallographic directions $a$, $b$ and $c$ may be used interchangeably with $x$, $y$ and $z$ respectively.}
	\label{tab:structurefactors}
	\centering
	\begin{ruledtabular}
	\begin{tabular}{c | c c c c | c c c}
				& \multicolumn{4}{c|}{$|S_R({\bf Q})|^2$}	& \multicolumn{3}{c}{$|P_i({\bf Q})|^2$}\\
	\multirow{2}{*}{$(H,K,L)$}			& $A$	& $G$	& $C$	& $F$ & $x$ & $y$ & $z$\\
		& $(\uparrow \downarrow \downarrow \uparrow)$ & $(\uparrow \downarrow \uparrow \downarrow)$ & $(\uparrow \uparrow \downarrow \downarrow)$	 & $(\uparrow \uparrow \uparrow \uparrow)$ & $a$	& $b$	& $c$ \\
	\hline
	$(3,0,1)$	&  0.07	&  0.22	& 11.73	&  3.98	& 0.34	& 1		& 0.66\\
	$(0,1,0)$	&  0  	&  0 	& 16 	&  0 	& 1		& 0		& 1\\
	$(1,0,0)$	& 15.51	&  0.49	&  0 	&  0 	& 0		& 1		& 1\\
	$(0,2,1)$	&  0 	& 15.71	&  0.29	&  0 	& 1		& 0.28	& 0.72\\
	$(0,1,2)$	&  0 	&  1.14	& 14.86	&  0 	& 1		& 0.86	& 0.14\\
	$(0,0,1)$	&  0 	& 15.71	&  0.29	&  0 	& 1		& 1		& 0\\
	\end{tabular}
	\end{ruledtabular}
\end{table}

\subsection{Spontaneous spin canting at zero field}

\begin{figure}[t!]
	\centering
	\includegraphics[width = \columnwidth]{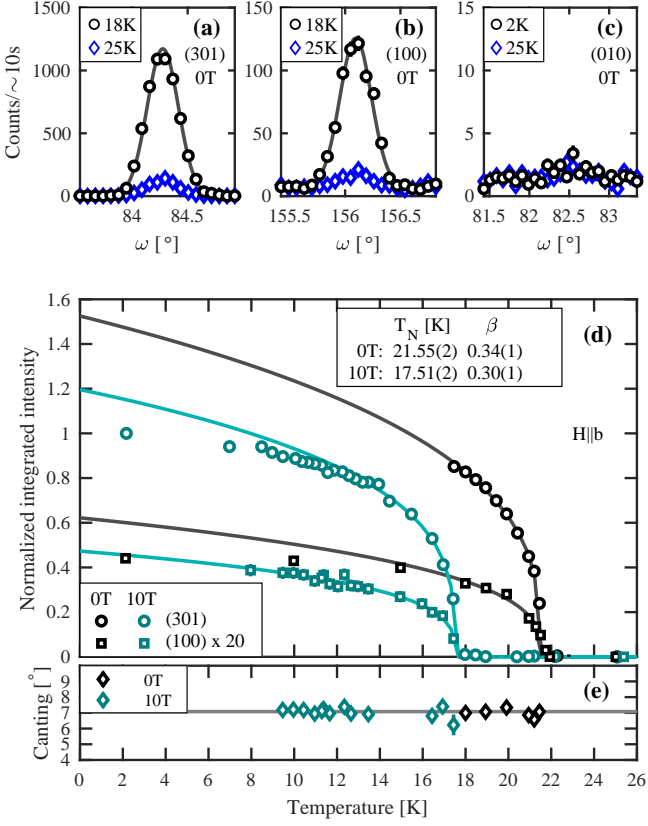}
	\caption{Neutron diffraction data from RITA-II. (a)-(c) rocking curves of $(3,0,1)$, $(1,0,0)$ and $(0,1,0)$ at $18$ or $2\,\mathrm{K}$ (black circles) and $25\,\mathrm{K}$(blue diamonds), all at $0\,\mathrm{T}$. The solid lines are Gaussian fits from which the integrated intensities are obtained. The canting angle of the zero-field structure is estimated from the intensity ratio between the magnetic contributions of $(3,0,1)$ and $(1,0,0)$. (d) Integrated magnetic intensity of $(3,0,1)$ (circles) and $(1,0,0)$ (squares) as a function of temperature at $0\,\mathrm{T}$ (black symbols) and $10\,\mathrm{T}$ along $b$ (green symbols). The intensities have been normalized to the value of $(3,0,1)$ at the lowest temperature and the intensity of $(1,0,0)$ has been multiplied by 20 for a better comparison of the temperature profiles. Backgrounds at $25\,\mathrm{K}$ have been subtracted. The solid lines show fits to a power law, $I \propto (T-T_N)^\beta$, for $T > 17\,\mathrm{K}$ at $0\,\mathrm{T}$ and $T > 13\,\mathrm{K}$ at $10\,\mathrm{T}$. The transition temperature, $T_N$, and critical exponent, $\beta$, were fitted collectively for the two peaks. (e) Spontaneous canting angle calculated from the intensity ratio of $(1,0,0)$ and $(3,0,1)$ for measurements done at $0\,\mathrm{T}$ (black symbols) and $10\,\mathrm{T}$ (green symbols). The horizontal line shows the value of the weighted mean of all data points, $\varphi = 7(1)^{\circ}$.}
	\label{fig:rita}
\end{figure}

In addition to the major $C_y$ spin component, a smaller symmetry component was identified by observation of magnetic intensity at the $(1,0,0)$ position. This peak mainly represents magnetic structures of $A$ symmetry with spins polarized along either $b$ or $c$. It is approximately one order of magnitude weaker than $(3,0,1)$ [compare Figs. \ref{fig:rita}(a) and \ref{fig:rita}(b)] which may be assumed to represent the major spin component when regarding the following argument: both $(3,0,1)$ and $(0,1,0)$ appear if a $C$ component is present but the two peaks represent different spin polarizations. $(3,0,1)$ is present for any spin orientation whereas $(0,1,0)$ is only present for components along $a$ or $c$. Since $(0,1,0)$ has no magnetic intensity [see Fig. \ref{fig:rita}(c)] we can exclude those two spin directions entirely. Hence, the $(3,0,1)$ magnetic Bragg peak may be assumed to solely represent a $C_y$ spin arrangement. 

\begin{figure}[t!]
	\centering
	\includegraphics[width = 0.9\columnwidth]{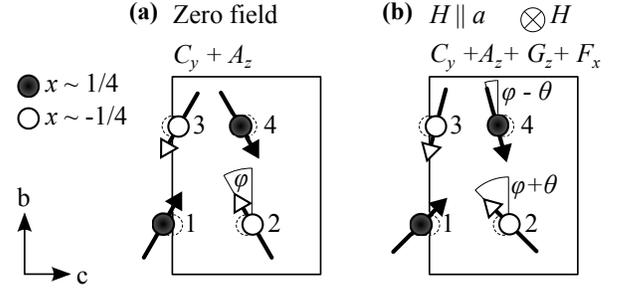}
	\caption{Projections in the $(b,c)$-plane of the magnetic structures of LiCoPO$_4$ at (a) zero field and for (b) $H \parallel a$. For clarity, only the four magnetic ions of the unit cell are shown and all angles are largely exaggerated. The ion positions deviate slightly from the high-symmetry positions (dashed circles). The applied field yields asymmetric total canting angles.}
	\label{fig:zero}
\end{figure}

Next, the basis vector corresponding to the $(1,0,0)$ Bragg peak is identified. The thermal average of the spin is most often maximized at low temperatures. Since an $A$ type component with spins along $b$ would produce spins of varying lengths, it is therefore reasonable to assume that the observed magnetic intensity at $(1,0,0)$ is instead due to a spin component along $c$. The result is a canting of the spins in the $(b,c)$-plane and the canting angle, $\varphi$, is estimated by comparing the intensity of $(1,0,0)$ with that of $(3,0,1)$. Following the above arguments, it is assumed that $(3,0,1)$ represents only a $C_y$ symmetry component and $(1,0,0)$ represents only an $A_z$ component such that the measured intensities may be written as in Eq. \eqref{eq:intensity}
\begin{align*}
	I_{(1,0,0)} & \propto \left|S_A^{(1,0,0)}\right|^2 \left|P_z^{(1,0,0)}\right|^2 f^2_{(1,0,0)},\\
	I_{(3,0,1)} & \propto \left|S_C^{(3,0,1)}\right|^2 \left|P_b^{(3,0,1)}\right|^2 f^2_{(3,0,1)},
\end{align*}
The spontaneous canting angle is then calculated from the corrected intensities, $I^{\mathrm{corr}}_{(1,0,0)}$ and $I^{\mathrm{corr}}_{(3,0,1)}$, as $\tan \varphi = \sqrt{I^{\mathrm{corr}}_{(1,0,0)}/I^{\mathrm{corr}}_{(3,0,1)}}$.
The usual Lorentz factor for two-axis diffractometers, $\sin (2 \theta)$, is also taken into account and although not entirely correct for the triple-axis setup \cite{pynn1975}, the correction is estimated to introduce an error of maximum 10\% for the two implicated Bragg peaks. The calculated angle is shown in Fig. \ref{fig:rita}(e) where both data at $0\,\mathrm{T}$ and $10\,\mathrm{T}$ along $b$ are shown. The canting angle is temperature independent below the transition temperature and it is also independent of the applied magnetic field. The magnetic structure is thus locked in with a spontaneous canting angle of $\varphi = 7(1)^{\circ}$ as estimated from a weighed mean of all data points in Fig. \ref{fig:rita}(e). The resulting zero-field structure is illustrated in Fig. \ref{fig:zero}(a). Note that the $(3,0,1)$ Bragg peak is relatively strong compared to $(1,0,0)$ and is therefore, to a larger extent, subject to extinction effects. Consequently, the calculated angle may be overestimated.

\begin{table}[b!]
	\caption{Irreducible representations, magnetic space groups and corresponding basis vectors for \textit{Pnma}.}
	\label{tab:irreps}
	\centering
	\begin{ruledtabular}
	\begin{tabular}{c c c c c c c c}
		$\Gamma_1$	& $\Gamma_2$	& $\Gamma_3$	& $\Gamma_4$	& $\Gamma_5$	& $\Gamma_6$	& $\Gamma_7$	& $\Gamma_8$\\
		\hline
		\footnotesize $Pnma$
	  & \footnotesize $Pnm'a'$
	  & \footnotesize $Pn'ma'$
	  & \footnotesize $Pn'm'a$
	  & \footnotesize $Pn'm'a'$
	  & \footnotesize $Pn'ma$
	  & \footnotesize $Pnm'a$
	  & \footnotesize $Pnma'$\\
		\hline
					& $F_x$		& 			& $G_x$		& $C_x$		&			& $A_x$		&\\
		$G_y$		&			& $F_y$		&			& 			& $A_y$		&			& $C_y$\\
					& $G_z$		&			& $F_z$		& $A_z$		& 			& $C_z$		&\\
	\end{tabular}
	\end{ruledtabular}
\end{table}

Both $(3,0,1)$  and $(1,0,0)$ appear at the same transition temperature -- see Fig. \ref{fig:rita}(d) -- and therefore reflect the same order parameter. Indeed, a power law with collectively fitted transition temperature, $T_N=21.55(2)\,\mathrm{K}$, and critical exponent, $\beta = 0.34(1)$, describe the recorded data well. However, note that the $C$ type structure polarized along $b$ and the $A$ type structure polarized along $b$ or $c$ are not contained within the same irreducible representation of the lithium orthophosphates, see Table \ref{tab:irreps}.

The Bragg peaks $(0,2,1)$, $(0,1,2)$ and $(0,0,1)$ also have magnetic intensity at $0\,\mathrm{T}$. These peaks are all present for a $C_y$ structure but may also represent a $G$ type component polarized along either $a$ or $b$, see Table \ref{tab:structurefactors}. A $G_y$ component is unlikely due to maximized moments at low temperatures and is not compatible with the observed ME effect, toroidal moment and weak ferromagnetism. Furthermore, $G_x$ is paired with $F_z$ in the irreducible representations, see Table \ref{tab:irreps}, and $F_z$ is not present \cite{kharchenko2003}. Therefore, the magnetic intensity at the $(0,2,1)$, $(0,1,2)$ and $(0,0,1)$ positions at $0\,\mathrm{T}$ may be subscribed to the major $C_y$ spin component. 

It is commented that the determined zero-field structure does not fully agree with earlier findings. A $C_z$ type rotation of the spins away from the $b$-axis was reported in Ref. \onlinecite{vaknin2002} based on the observation of the $(0,1,0)$ magnetic peak. However, as seen in Fig. \ref{fig:rita}(c), we observe zero magnetic intensity at the $(0,1,0)$ position. A maximum of the rotation angle of $0.7(3)^{\circ}$ is estimated from the error on the measured zero intensity. This is contrasted by the $4.6^{\circ}$ reported in Ref. \onlinecite{vaknin2002}.
%At present, we have no explanation for the discrepancy.
One possible explanation for the discrepancy might be found in slightly different levels of Li in different samples. Previously, changes in atomic bond lengths and magnetic properties of Li$_z$CoPO$_4$ with $z = 0.2,0.7$ as compared to the stoichiometric compound, LiCoPO$_4$, were reported \cite{ehrenberg2009}. It is conceivable that small variations in Li contents between samples may bring about small differences in the exact magnetic structure. As already mentioned, our sample was found to have a Li occupancy of $1.03(5)$.

It has been repeatedly suggested \cite{vaknin2002,vanAken2008,zimmermann2009} that the zero-field structure of LiCoPO$_4$ has lower symmetry than the originally proposed magnetic point group $mmm'$ \cite{santoro1966}. The observed $4.6^{\circ}$ rotation of spins restricts symmetry to $2_x'/m_x$ which is further reduced to $2'_x$ when requiring a weak ferromagnetic component along $b$. Indeed, optical second harmonic generation measurements advocate that the point group symmetry is $2'_x$ \cite{zimmermann2009}. This point group allows for a toroidal moment \cite{schmid2008} and the linear ME effect with tensor elements $\alpha_{ab}, \alpha_{ba} \neq 0$ \cite{rivera2009}, consistent with measurements \cite{rivera1994}. In addition, $2'_x$ allows the tensor elements $\alpha_{ac}, \alpha_{ca} \neq 0$ which are not measurably different from zero \cite{rivera1994} but the spin rotation angle introduces only a small deviation from $mmm'$. Furthermore, as the point group merely yields the \textit{allowed} ME tensor elements they are not \textit{necessarily} active.

Thus neutron diffraction \cite{vaknin2002}, SQUID \cite{kharchenko2001} and optical second harmonic generation measurements \cite{zimmermann2009,vanAken2008} all paint a picture of LiCoPO$_4$ having magnetic point group $2'_x$ in its zero-field state. In contrast, our observation of a spontaneous spin canting rather than a rotation leads to the magnetic point group $2_z/m'_z$. This point group also allows for a toroidal moment and the ME tensor elements $\alpha_{aa},\alpha_{ab},\alpha_{ba},\alpha_{bb},\alpha_{cc} \neq 0$ where only the off-diagonal elements are measurably different from zero. Again, we note that the canting angle only presents a small deviation from $mmm'$. $2_z/m'_z$ does not support a ferromagnetic moment along $b$ rendering it inconsistent with observations. However, removing the twofold axis enables a ferromagnetic moment in the $(a,b)$-plane. Thus, the magnetic point group $m_z'$ is consistent with our neutron diffraction data and a weak ferromagnetic moment along $b$. Note, however, that it is not consistent with the observed optical second harmonic generation signal \cite{zimmermann2009,vanAken2008}.

Interestingly, $m'_z$ is also consistent with the previous neutron diffraction study when using a different -- but still correct -- interpretation of the presented data. The rotation of the spins towards $c$ was established based on observation of the $(0,1,0)$ magnetic Bragg peak. However, this rotation might equally well be towards $a$. Assuming such a rotation results in magnetic point group $2_z/m'_z$ which again needs relaxing to $m'_z$ to allow for a ferromagnetic moment along $b$. In addition, the $C_x$ component belongs to the same irreducible representation as the $A_z$ component [see Table \ref{tab:irreps}] and as is deducted in the next section; the two components combined yield a favorable energy term via the DM interaction. Therefore, our observations may in fact be consistent with the previous studies and the magnetic point group of the zero-field structure of LiCoPO$_4$ is $m'_z$.

\subsection{Field-induced spin canting for H$\parallel$a}

\begin{figure*}
	\centering
	\includegraphics[width = \textwidth]{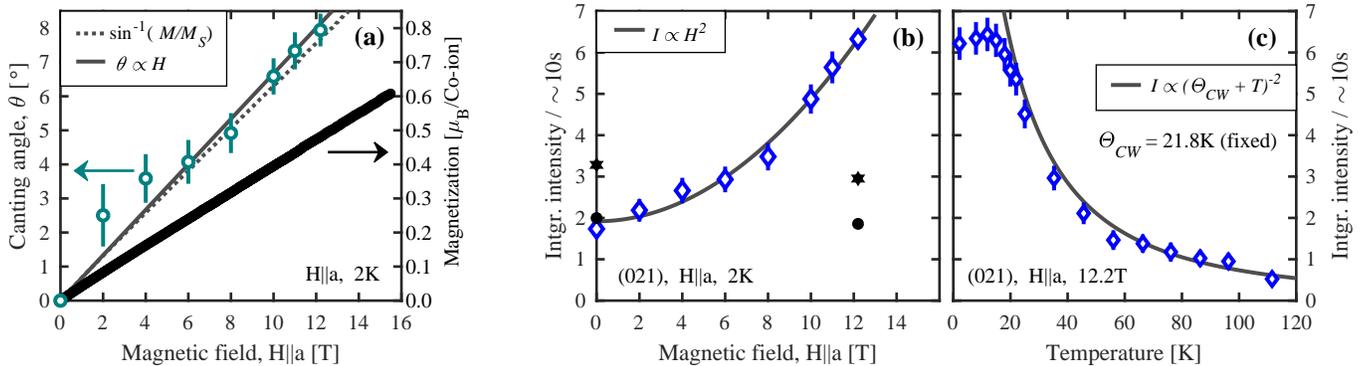}
	\caption{Magnetization, field-induced canting angle and integrated neutron intensity of $(0,2,1)$ for magnetic fields applied along $a$. (a) magnetization (thick black line) and field-induced canting angle (green circles) calculated from the neutron intensity of $(0,2,1)$ as a function of applied field. Both are to a good approximation linearly proportional to the field strength. The solid line shows a linear fit to the canting angle whereas the dashed line shows the angle as calculated from the magnetization with $M_s = 3.6\,\mu_B$/ion. (b) and (c) integrated intensity (blue diamonds) of $(0,2,1)$ as a function of applied field at $2\,\mathrm{K}$ and as a function of temperature at $12.2\,\mathrm{T}$ respectively. The field dependence in (b) and the temperature dependence in (c) have been fitted to a quadratic and a Curie-Weiss law squared respectively (solid lines). The black symbols in (b) show intensities for $(0,1,2)$ (circles) and $(0,0,1)$ (stars) at $0\,\mathrm{T}$ and $12.2\,\mathrm{T}$. Note that the intensities for these peaks are scaled to appear together with the intensity of $(0,2,1)$ in order to demonstrate that they show no or only little field dependence.}
	\label{fig:a-axis}
\end{figure*}

%For magnetic fields applied along the easy $b$-axis, LiCoPO$_4$ displays 3 phase transitions before saturation at $\sim28\,\mathrm{T}$ \cite{fogh2017,kharchenko2010}, the first of which occurs at $\sim12\,\mathrm{T}$. As expected, a different behavior is found for magnetic fields applied along the harder axis $a$. Here no magnetic phase transitions were detected for fields probed up to $16\,\mathrm{T}$.

For magnetic fields applied along $a$, LiCoPO$_4$ is linearly magnetized with the field as seen in the magnetization data in Fig. \ref{fig:a-axis}(a). A ferromagnetic contribution to the spin structure is induced with $S^a = \alpha H$ and fitted slope $\alpha = 0.0395(1)\,\mu_B/\mathrm{T}$. Furthermore, yet another antiferromagnetic component exists in addition to the established main structure of $C_y$ symmetry and the minor $A_z$ component. This extra component is manifested by an increase in the intensity of the $(0,2,1)$ magnetic Bragg peak as a function of applied field, see Fig. \ref{fig:a-axis}(b). The magnetic origin of the $(0,2,1)$ intensity is confirmed by its temperature dependence which follows a Curie-Weiss law squared, see Fig. \ref{fig:a-axis}(c).

The $(0,2,1)$ peak represents mainly spin arrangements of symmetry $G$ and to a smaller extent structures of symmetry $C$, \textit{cf.} Table \ref{tab:structurefactors}. All spin orientations are possible and more information is therefore needed in order to pin down which magnetic structure the additional intensity of $(0,2,1)$ signifies. Again, the argument follows a process of elimination using two other magnetic Bragg peaks: $(0,1,2)$ and $(0,0,1)$.

The $(0,1,2)$ peak is present for any $C$ spin structures. This peak has no additional field-induced intensity [see Fig. \ref{fig:a-axis}(b)] and consequently any additional $C$ spin elements are ruled out. Finally, $(0,0,1)$ represents $G$ symmetry with spins polarized along $a$ or $b$. Again, this peak shows no change upon applying a magnetic field along $a$ [see Fig. \ref{fig:a-axis}(b)] and these magnetic structure types may too be rejected. The only remaining possible magnetic structure as a contributor to the $(0,2,1)$ field-induced intensity is then $G_z$. This component comes as an addition to the already established major $C_y$ component and the smaller $A_z$ component. An asymmetry is introduced in the canting angles such that spins $(1,2)$ and $(3,4)$ form pairs with canting angles $\phi+\theta$ and $\phi-\theta$ respectively. Here $\theta \equiv \theta(H)$ is the field-induced canting angle. The resulting magnetic structure for magnetic fields applied along $a$ is shown in Fig. \ref{fig:zero}(b).

The size of $\theta$ is now estimated. As previously argued, it may be assumed that at $0\,\mathrm{T}$, $(0,2,1)$ only reflects the $C_y$ structure. Any additional intensity upon applying a field then originates from the $G_z$ component:
\[
	I_{(0,2,1)}(H) - I_{(0,2,1)}(0\,\mathrm{T}) \propto \left|S_{G}^{(0,2,1)}\right|^2 \left|P_z^{(0,2,1)}\right|^2.
\]
This is to be compared to the intensity of $(0,2,1)$ at $0\,\mathrm{T}$:
\[
	I_{(0,2,1)}(0\,\mathrm{T}) \propto \left|S_{C}^{(0,2,1)}\right|^2 \left|P_y^{(0,2,1)}\right|^2.
\]
Since only one peak is involved in the determination of the field-induced canting angle there is no need to correct for the magnetic form factor or Lorentz factor and any extinction or absorption effects may be neglected. The field-induced canting angle is then calculated as $\tan \theta = \sqrt{ \frac{I^{\mathrm{corr}}_{(0,2,1)}(H) - I^{\mathrm{corr}}_{(0,2,1)}(0\,\mathrm{T}) }{I^{\mathrm{corr}}_{(0,2,1)}(0\,\mathrm{T})}}$ and is to a good approximation linear as a function of applied field along $a$: $\theta = \beta H$ with fitted slope $\beta = 0.012(1)\,\mathrm{rad/T}$ [see Fig. \ref{fig:a-axis}(a)]. The field-induced canting angle as deduced from the magnetization, $\sin \theta = M/M_S$, is also shown in Fig. \ref{fig:a-axis}(a) and substantiates the link between $F_x$ and $G_z$. Furthermore,
%The magnetic field-induced intensity of the $(0,2,1)$ peak suggests a connection between the $G_z$ order and the ferromagnetic component $F_x$.
since the neutron intensity is proportional to the ordered magnetic moment squared, a linear coupling between the ferromagnetic moment and canted moment would result in a quadratic increase in the neutron intensity of $(0,2,1)$ as a function of applied field. This is indeed the case as shown in Fig. \ref{fig:a-axis}(b). Here the solid line is a fit to a quadratic dependence, $I \propto H^2$. The measured intensity is clearly well described by the fit. Additionally,  the symmetry elements $G_z$ and $F_x$ belong to the same irreducible representation, see Table \ref{tab:irreps}.%, further supporting the hypothesis that the two components are linked. 

\subsection{Dzyaloshinskii-Moriya interaction}

An estimate of the size of the DM interaction in LiCoPO$_4$ may be obtained from the field-induced spin canting. A similar calculation was previously performed for the sister compound LiNiPO$_4$ and the analysis in Ref. \onlinecite{jensen2009_2} is directly applicable here. Symmetry arguments lead to the only allowed DM coefficients ${\bf D}_{14} = (0,D_{14}^b,0) = - {\bf D}_{23}$ and ${\bf D}_{12} = (0,D_{12}^b,0) = {\bf D}_{34}$. These yield terms in the Hamiltonian of the form:
\begin{align*}
	\mathcal{H}_{\mathrm{DM}}^1  & = {\bf D}_{14} \cdot \left( {\bf S}_1 \times {\bf S}_4 \right) - {\bf D}_{14} \cdot \left( {\bf S}_2 \times {\bf S}_3 \right)\\
	 							& = D_{14}^b \left( S_1^c S_4^a - S_1^a S_4^c - S_2^c S_3^a + S_2^a S_3^c \right) \quad \mathrm{and}\\
	 \mathcal{H}_{\mathrm{DM}}^2 & = {\bf D}_{12} \cdot \left( {\bf S}_1 \times {\bf S}_2 \right) + {\bf D}_{12} \cdot \left( {\bf S}_3 \times {\bf S}_4 \right)\\
	 							& = D_{12}^b \left( S_1^c S_2^a - S_1^a S_2^c + S_3^c S_4^a - S_3^a S_4^c \right).
\end{align*}
The spin component along $a$ is finite for $H \parallel a$ and assumed equal at all sites, i.e. $S_1^a = S_2^a = S_3^a = S_4^a = S^a > 0$. In this case, both terms favor a $G_z$ type order and this is exactly what we observe. The ferromagnetic moment along $a$ therefore induces -- via the DM interaction -- an antiferromagnetic spin component of symmetry $G_z$.

The field-induced $G_z$ component leaves the nearest neighbor spin pairs $(1,4)$ and $(2,3)$ antiparallel and hence no energy change is to be expected from the term $\mathcal{H}^1_{\mathrm{DM}}$ nor from the nearest neighbor exchange term. On the other hand, the term $\mathcal{H}^2_{\mathrm{DM}}$ does indeed yield a finite energy contribution for a $G_z$ component. The strength of the DM interaction may be estimated by balancing the different energy contributions for spins deviating from the easy axis, $b$:
\begin{align*}
\left.
	\begin{matrix}
		\mathcal{H}_{\mathrm{DM}} = 4 D^b_{12} S^a S \sin \theta \\
		\mathcal{H}_{\mathrm{ani}} = 4 \mathfrak{D}^c S^2 \sin^2 \theta	
	\end{matrix}
	\right\rbrace \Rightarrow \frac{D^b_{12}}{\mathfrak{D}^c} = \frac{-S \sin \theta}{S^a} \approx -S\frac{\theta}{S^a},
\end{align*}
where $\mathfrak{D}^c$ is the single-ion anisotropy constant for spin components along $c$, $S=3.6\,\mathrm{\mu_B}$ the saturated moment, $\sin \theta \approx \theta$ holds for small canting angles, $\theta = \beta H$ and $S^a = \alpha H$. With the fitted coefficients $\beta = 0.012(1)\,\mathrm{rad/T}$ and $\alpha = 0.0395(1)\,\mathrm{\mu_B/T}$ the ratio becomes $D^b_{12} / \mathfrak{D}^c \approx -1.1$. Note that this is an upper bound for the size of the DM interaction as the above simple calculation neglects any competing exchange interactions which may also influence the spin canting.

Thus, the DM interaction in LiCoPO$_4$ may be as large as the single-ion anisotropy along $c$. The full spin Hamiltonian of LiCoPO$_4$ has not been determined yet, but limited inelastic neutron scattering data shows an almost dispersionless spin excitation along the $(0,K,0)$ direction and a single-ion anisotropy constant of $\mathfrak{D}^c \approx 0.7\,\mathrm{meV}$ is suggested \cite{vaknin2002,tian2008,Note1}. This is a very strong DM interaction and its possible role as a generator for the ME effect in LiCoPO$_4$ is discussed in the following.

Magnetostrictive mechanisms successfully explain the ME effect in LiNiPO$_4$ \cite{jensen2009_2,toftpetersen2017} and LiFePO$_4$ \cite{toftpetersen2015} based on magnetic field-induced changes in the exchange and DM interactions respectively. A similar model would be expected to describe the effect in LiCoPO$_4$. However, so far a satisfactory model has eluded all our efforts -- both when considering magnetic field-induced changes in the exchange and DM interactions individually and combined. Such microscopic models inherently result in a ME coefficient, $\alpha_{\parallel}$, proportional to $\chi_{\parallel} \langle S \rangle^2$, i.e. the magnetic susceptibility and the order parameter. The susceptibility drops at low temperatures in a collinear antiferromagnet and the order parameter levels out after the initial increase at the transition. Hence the temperature dependence of $\alpha_{\parallel}$ has a maximum below the transition as seen in LiMnPO$_4$, LiNiPO$_4$ and LiFePO$_4$ [re-visit Fig. \ref{fig:ME}]. However, for LiCoPO$_4$ $\alpha_{\parallel}$ does not display such maximum as a function of temperature. In fact, its temperature profile resembles that of the order parameter and the curves are similar for $\alpha_{\parallel}$ and $\alpha_{\perp}$.

As discussed in the introduction, it was previously proposed that the spin-orbit coupling is a central element in fully understanding the ME effect in the lithium orthophosphates. However, \textit{ab initio} calculations considering both spin and orbital momentum on equal footing still fail to correctly predict the size of $\alpha_{\parallel}$ for $T \rightarrow 0$ in LiFePO$_4$ \cite{scaramucci2012}. The spin-orbit coupling is expected to be larger in the sister compound, LiCoPO$_4$ and similar first-principle computations may be expected to produce larger ME coefficients. To our best knowledge such calculations have yet to be performed. Nevertheless, our neutron diffraction data show that there is indeed a large DM interaction in LiCoPO$_4$ which in turn relates to the spin-orbit coupling. Therefore, it remains that the spin-orbit coupling plays an important role in generating the ME effect in LiCoPO$_4$ -- and most likely in the entire family of compounds. This emphasizes the need for more theoretical work and improved \textit{ab initio} calculations in order to elucidate the missing mechanism(s) governing the linear ME effect in LiCoPO$_4$ and even better; explain the link between the spin-orbit coupling and the ME effect in the lithium orthophosphates in general. Moreover, spin excitation measurements would enable modeling of the spin Hamiltonian of LiCoPO$_4$ and thereby provide a better understanding of the magnetic interactions in the system.

\section{Conclusions}

Intricate details of the zero-field magnetic structure of LiCoPO$_4$ were investigated in hope of illuminating the microscopic mechanism behind the large magnetoelectric effect in LiCoPO$_4$. The Co$^{2+}$ ions mainly order in a commensurate antiferromagnetic structure of $C_y$ symmetry. Additionally, we discover a spontaneous spin canting of $\varphi = 7(1)^{\circ}$ originating in an $A_z$ spin component. The resulting zero-field magnetic structure belongs to the magnetic point group $m'_z$, consistent with previously reported experimental results.

For magnetic fields applied along $a$, a second minor spin component of symmetry $G_z$ is induced. The canting angle increases to a good approximation linearly with the applied field and is shown to be induced via the Dzyaloshinskii-Moriya interaction by the ferromagnetic moment along $a$. The upper limit for the size of the Dzyaloshinskii-Moriya interaction was estimated to be approximately equal to that of the single-ion anisotropy constant along $c$. This shows that the spin-orbit coupling is strong in LiCoPO$_4$ and we discuss how it may be linked to its large magnetoelectric effect.

\section*{Acknowledgements}

Work was supported by the Danish Agency for Science and Higher Education under DANSCATT.
Neutron diffraction experiments were performed at the Swiss spallation neutron source SINQ, Paul Scherrer Institute, Villigen, Switzerland.
Ames Laboratory is operated by the U.S. Department of Energy by Iowa State University under Contract No. DE-AC02-07CH11358.

%\bibliographystyle{unsrt}

%\bibliography{../../bibbase}

\begin{thebibliography}{10}

\bibitem{landau_lifshitz}
L.~D. Landau and E.~M. Lifshitz.
\newblock {\em Electrodynamics of Continuous Media}.
\newblock Pergamon, Oxford, UK, 1960.

\bibitem{dzyaloshinskii1959}
I.~E. Dzyaloshinskii.
\newblock On the magneto-electrical effect in antiferromagnets.
\newblock {\em J. Exp. Theor. Phys.} {\bf 10}, 628, (1960).

\bibitem{astrov1960}
D.~N. Astrov.
\newblock The magnetoelectric effect in antiferromagnetics.
\newblock {\em J. Exp. Theor. Phys.} {\bf 11}, 708, (1960).

\bibitem{astrov1961}
D.~N. Astrov.
\newblock Magnetoelectric effect in chromium oxide.
\newblock {\em J. Exp. Theor. Phys.} {\bf 13}, 729, (1961).

\bibitem{sergienko2006}
I.~A. Sergienko and E.~Dagotto.
\newblock Role of the dzyaloshinskii-moriya interaction in multiferroic
  perovskites.
\newblock {\em Phys. Rev. B} {\bf 73}, 094434 (2006).

\bibitem{kenzelmann2006}
M.~Kenzelmann, A.~B. Harris, A.~Aharony, O.~Entin-Wohlman abd T.~Yildirim,
  Q.~Huang, S.~Park, G.~Lawes, C.~Broholm, N.~Rogado, R.~J. Cava, K.~H. Kim,
  G.~Jorge, and A.~P. Ramirez.
\newblock Field dependence of magnetic ordering in kagomé-staircase compound
  {Ni$_3$V$_2$O$_8$}.
\newblock {\em Phys. Rev. B} {\bf 74}, 014429 (2006).

\bibitem{kimura2006}
T.~Kimura, J.~C. Lashley, and A.~P. Ramirez.
\newblock Inversion-symmetry breaking in the noncollinear magnetic phase of the
  triangular-lattice antiferromagnet {CuFeO$_2$}.
\newblock {\em Phys. Rev. B} {\bf 73}, 220401 (2006).

\bibitem{kimura2007}
T.~Kimura.
\newblock Spiral magnets as magnetoelectrics.
\newblock {\em Annu. Rev. Mater. Res.} {\bf 37}, 387-413 (2007).

\bibitem{spaldin2008}
N.~A. Spaldin, M.~Fiebig, and M.~Mostovoy.
\newblock The toroidal moment in condensed-matter physics and its relation to
  the magnetoelectric effect.
\newblock {\em J. Phys. Cond. Matt.} {\bf 20}, 434203 (2008).

\bibitem{toftpetersen2017}
R.~Toft-Petersen, E.~Fogh, T.~Kihara, J.~Jensen, K.~Fritsch, J.~Lee, G.~E.
  Granroth, M.~B. Stone, D.~Vaknin, H.~Nojiri, and N.~B. Christensen.
\newblock Field-induced reentrant magnetoelectric phase in {LiNiPO$_4$}.
\newblock {\em Phys. Rev. B} {\bf 95}, 064421 (2017).

\bibitem{khrustalyov2016}
V.~M. Khrustalyov, V.~M. Savytsky, and M.~F. Kharchenko.
\newblock Magnetoelectric effect in antiferromagnetic {LiCoPO$_4$} in pulsed
  magnetic fields.
\newblock {\em Low Temp. Phys.} {\bf 42}, 280-285 (2016).

\bibitem{fogh2017}
E.~Fogh, R.~Toft-Petersen, E.~Ressouche, C.~Niedermayer, S.~L. Holm,
  M.~Bartkowiak, O.~Prokhnenko, S.~Sloth, F.~W. Isaksen, D.~Vaknin, and N.~B.
  Christensen.
\newblock Magnetic order, hysteresis, and phase coexistence in magnetoelectric
  licopo$_4$.
\newblock {\em Phys. Rev. B} {\bf 96}, 104420 (2017).

\bibitem{iniguez2008}
J.~Ìñiguez.
\newblock First-principles approach to lattice-mediated magnetoelectric
  effects.
\newblock {\em Phys. Rev. Lett.} {\bf 101}, 117201 (2008).

\bibitem{mostovoy2010}
M.~Mostovoy, A.~Scaramucci, N.~A. Spaldin, and K.~T. Delaney.
\newblock Temperature-dependent magnetoelectric effect from first principles.
\newblock {\em Phys. Rev. Lett.} {\bf 105}, 087202 (2010).

\bibitem{malashevich2012}
A.~Malashevich, S.~Coh, I.~Souza, and D.~Vanderbilt.
\newblock Full magnetoelectric response of {Cr$_2$O$_3$} from first principles.
\newblock {\em Phys. Rev. B} {\bf 86}, 094430 (2012).

\bibitem{mu2014}
S.~Mu, A.~L. Wysocki, and K.~D. Belashchenko.
\newblock First-principles microscopic model of exchange-driven magnetoelectric
  response with application to {Cr$_2$O$_3$}.
\newblock {\em Phys. Rev. B} {\bf 89}, 174413 (2014).

\bibitem{scaramucci2012}
A.~Scaramucci, E.~Bousquet, M.~Fechner, M.~Mostovoy, and N.~A. Spaldin.
\newblock Linear magnetoelectric effect by orbital magnetism.
\newblock {\em Phys. Rev. Lett.} {\bf 109}, 197203 (2012).

\bibitem{rado1961}
G.~T. Rado and V.~J. Folen.
\newblock Observation of the magnetically induced magnetoelectric effect and
  evidence for antiferromagnetic domains.
\newblock {\em Phys. Rev. Lett.} {\bf 7}, 310-311 (1961).

\bibitem{mercier}
M.~Mercier.
\newblock {\em Étude de l'effet magnetoelectrique sur de composés de type
  olivine, perovskite et grenat}.
\newblock PhD thesis, Université de Grenoble, September 1969.

\bibitem{toftpetersen2015}
R.~Toft-Petersen, M.~Reehuis, T.~B.~S. Jensen, N.~H. Andersen, J.~Li, M.~Duc
  Le, M.~Laver, C.~Niedermayer, B.~Klemke, K.~Lefmann, and D.~Vaknin.
\newblock Anomalous magnetic structure and spin dynamics in magnetoelectric
  {LiFePO$_4$}.
\newblock {\em Phys. Rev. B} {\bf 92}, 024404 (2015).

\bibitem{rivera1994}
J.-P. Rivera.
\newblock The linear magnetoelectric effect in {LiCoPO$_4$} revisited.
\newblock {\em Ferroelectrics} \textbf{161}, 147-164 (1994).

\bibitem{wieglhofer}
W.~S. Wieglhofer and A.~Lakhtakia.
\newblock {\em Introduction to Complex Mediums for Optics and
  Electromagnetics}.
\newblock SPIE, Bellingham, Washington, USA, 2003.

\bibitem{newnham1965}
R.~E. Newnham and M.~J. Redman.
\newblock Crystallographic data for {LiMgPO$_4$}, {LiCoPO$_4$} and
  {LiNiPO$_4$}.
\newblock {\em J. Am. Ceram. Soc.} {\bf 48}, 547 (1965).

\bibitem{kubel1994}
F.~Kubel.
\newblock Crystal structure of lithium cobalt double orthophosphate,
  {LiCoPO$_4$}.
\newblock {\em Zeitschrift für Kristallographie} {\bf 209}, 755 (1994).

\bibitem{santoro1966}
R.~P. Santoro, D.~J. Segal, and R.~E. Newnham.
\newblock Magnetic properties of {LiCoPO$_4$} and {LiNiPO$_4$}.
\newblock {\em J. Phys. Chem. Solids} {\bf 27}, 1192-1193 (1966).

\bibitem{szewczyk2011}
A.~Szewczyk, M.~U. Gutowska, J.~Wieckowski, A.~Wisniewski, R.~Puzniak,
  R.~Diduszko, Yu. Kharchenko, M.~F. Kharchenko, and H.~Schmid.
\newblock Phase transitions in single-crystalline magnetoelectric {LiCoPO$_4$}.
\newblock {\em Phys. Rev. B} {\bf 84}, 104419 (2011).

\bibitem{vaknin2002}
D.~Vaknin, J.~L. Zarestky, L.~L. Miller, J.-P. Rivera, and H.~Schmid.
\newblock Weakly coupled antiferromagnetic planes in single-crystal
  {LiCoPO$_4$}.
\newblock {\em Phys. Rev. B} {\bf 65}, 224414 (2002).

\bibitem{kharchenko2010}
N.~F. Kharchenko, V.~M. Khrustalev, and V.~N. Savitskii.
\newblock Magnetic field induced spin reorientation in the strongly anisotropic
  antiferromagnetic crystal {LiCoPO$_4$}.
\newblock {\em Low Temp. Phys.} {\bf 36}, 558-564 (2010).

\bibitem{kharchenko2003}
Yu. Kharchenko, N.~Kharchenko, M.~Baran, and R.~Szymczak.
\newblock Weak ferromagnetism in magnetoelectrics {LiCoPO$_4$} and
  {LiNiPO$_4$}.
\newblock {\em Magnetoelectric Interaction Phenomena in Crystals}, 227-234, 2004.

\bibitem{vanAken2008}
B.~B.~Van Aken, J.-P. Rivera, H.~Schmid, and M.~Fiebig.
\newblock Anisotropy of antiferromagnetic {180$^{\circ}$} domains in
  {LiCoPO$_4$} and {LiNiPO$_4$}.
\newblock {\em Phys. Rev. Lett.} {\bf 101}, 157202 (2008).

\bibitem{ederer2007}
C.~Ederer and N.~A. Spaldin.
\newblock Towards a microscopic theory of toroidal moments in bulk periodic
  crystals.
\newblock {\em Phys. Rev. B} {\bf 76}, 214404 (2007).

\bibitem{vanAken2007}
B.~B.~Van Aken, J.-P. Rivera, H.~Schmid, and M.~Fiebig.
\newblock Observation of ferrotoroidic domains.
\newblock {\em Nature Letters} {\bf 449}, 702-705 (2007).

\bibitem{zimmermann2014}
A.~S. Zimmermann, D.~Meier, and M.~Fiebig.
\newblock Ferroic nature of magnetic toroidal order.
\newblock {\em Nature Communications} {\bf 5}, 4796 (2014).

\bibitem{khrustalyov2017}
V.~M. Khrustalyov, V.~M. Savytsky, and M.~F. Kharchenko.
\newblock {$(H,T_i)$}-diagram of magnetic transformations induced by a pulsed
  magnetic field in antiferromagnetic {LiCoPO$_4$}.
\newblock {\em Low Temp. Phys.} {\bf 43}, 1332-1337 (2017).

\bibitem{rodriguez-carvajal1993}
J.~Rodriguez-Carvajal.
\newblock Recent advances in magnetic structure determination by neutron powder
  diffraction.
\newblock {\em Physica B}, {\bf 192}, 55-69 (1993).

\bibitem{pynn1975}
R.~Pynn.
\newblock Lorentz factors for triple-axis spectrometers.
\newblock {\em Acta Cryst.} {\bf B31}, 2555-2556, (1975).

\bibitem{ehrenberg2009}
H.~Ehrenberg, N.~N. Bramnik, A.~Senyshyn, and H.~Fuess.
\newblock Crystal and magnetic structures of electrochemically delithiated
  {Li$_{1-x}$CoPO$_4$} phases.
\newblock {\em Solid State Sciences} {\bf 11}, 18-23 (2009).

\bibitem{zimmermann2009}
A.~S. Zimmermann, B.~B.~Van Aken, H.~Schmid, J.-P. Rivera, J.~Li, D.~Vaknin,
  and M.~Fiebig.
\newblock Anisotropy of antiferromagnetic {180$^{\circ}$} domains in
  magnetoelectric {Li$M$PO$_4$} ({$M$ = Fe, Co, Ni}).
\newblock {\em Eur. Phys. J. B} {\bf 71}, 355-360 (2009).

\bibitem{schmid2008}
H.~Schmid.
\newblock Some symmetry aspects of ferroics and single phase multiferroics.
\newblock {\em J. Phys. Condens. Matter} {\bf 20}, 434201 (2008).

\bibitem{rivera2009}
J.-P. Rivera.
\newblock A short review of the magnetoelectric effect and related experimental
  techniques on single phase (multi-) ferroics.
\newblock {\em eur. Phys. J. B} \textbf{71}, 299-313 (2009).

\bibitem{kharchenko2001}
N.~F. Kharchenko, Yu.~N. Kharchenko, R.~Szymczak, M.~Baran, and H.~Schmid.
\newblock Weak ferromagnetism in the antiferromagnetic magnetoelectric crystal
  {LiCoPO$_4$}.
\newblock {\em Low Temp. Phys.} \textbf{27}, 895 (2001).

\bibitem{jensen2009_2}
T.~B.~S. Jensen, N.~B. Christensen, M.~Kenzelmann, H.~M. Rønnow,
  C.~Niedermayer, N.~H. Andersen, K.~Lefmann, J.~Schefer, M.~Zimmermann, J.~Li,
  J.~L. Zarestky, and D.~Vaknin.
\newblock Field-induced magnetic phases and electric polarization in
  {LiNiPO$_4$}.
\newblock {\em Phys. Rev. B} {\bf 79}, 092412 (2009).

\bibitem{tian2008}
W.~Tian, J.~Li, J.~W. Lynn, J.~L. Zarestky, and D.~Vaknin.
\newblock Spin dynamics in the magnetoelectric effect compound {LiCoPO$_4$}.
\newblock {\em Phys. Rev. B} {\bf 78}, 184429 (2008).

\bibitem{Note1}
We have examined crystals from the same batch as that used in Ref. [43] and find a significantly lower transition temperature, $T_N = 17.3(1)\protect  \tmspace +\thinmuskip {.1667em}\protect \mathrm {K}$. Furthermore, a Rietveld refinement of our neutron diffraction data yields satisfactory results exclusively when introducing Ni as well as Co on the magnetic site. Hence, our results suggest that the crystals may not be pure LiCoPO$_4$, but possibly Ni-doped from a crucible-growth.

\end{thebibliography}

\end{document}